\pdfoutput=1

\documentclass[12pt, a4paper]{article}

\usepackage{amsmath}
\usepackage{amsfonts}
\usepackage{amssymb}
\usepackage{graphicx, rotating}
\usepackage{epsfig}
\usepackage{latexsym}
\usepackage{graphicx}
\usepackage{color}
\usepackage{amsmath,bm,amssymb}
\usepackage{cite}
\usepackage{slashed}
\usepackage{hyperref}
\hypersetup{colorlinks=true, citecolor=bluscuro, linkcolor=black, urlcolor=bluscuro}
\definecolor{rossos}{cmyk}{0,1,1,0.55}
\definecolor{bluscuro}{rgb}{0.15, 0.2, .85}
\definecolor{bluchiaro}{cmyk}{1,.3,0.,0.1}
\usepackage{amsmath}
\numberwithin{equation}{section}
\def\ee{\end{equation}}
\def\ba{\begin{eqnarray}}
\def\ea{\end{eqnarray}}

\def\bq{\begin{quote}}
\def\eq{\end{quote}}

 at 10truept

\newcommand{\beq}{\begin{equation}}
\newcommand{\eeq}{\end{equation}}
\newcommand{\beqa}{\begin{eqnarray}}
\newcommand{\eeqa}{\end{eqnarray}}
\newcommand{\bea}{\begin{eqnarray}}
\newcommand{\eea}{\end{eqnarray}}
\newcommand{\p}{\partial}
\newcommand{\al}{\alpha}
 \newcommand{\be}{\beta}

\def\ltap{\ \raise.3ex\hbox{$<$\kern-.75em\lower1ex\hbox{$\sim$}}\ }
\def\gtap{\ \raise.3ex\hbox{$>$\kern-.75em\lower1ex\hbox{$\sim$}}\ }
\def\gl{\ \raise.5ex\hbox{$>$}\kern-.8em\lower.5ex\hbox{$<$}\ }
\def\roughly#1{\raise.3ex\hbox{$#1$\kern-.75em\lower1ex\hbox{$\sim$}}}

\def\t{\tau}
\def\bx{\vec{x}}
\def\del{\partial}
\def\bnabla{\vec{\nabla}}
\def\d{{\rm d}}
\newcommand{\lsim}{\,\raisebox{-.1ex}{$_{\textstyle <}\atop^{\textstyle\sim}$}\,}
\newcommand{\gsim}{\,\raisebox{-.3ex}{$_{\textstyle >}\atop^{\textstyle\sim}$}\,}

\newcommand{\arXiv}[2]{\href{http://arxiv.org/pdf/#1}{{\tt [#2/#1]}}}
\newcommand{\arXivold}[1]{\href{http://arxiv.org/pdf/#1}{{\tt [#1]}}}

 \def\al{\alpha}

\begin{document}

%FRONTPAGE2%%%%%%
\begin{titlepage}
\begin{flushright}
%DESY 16-xxx\\
%CERN ...
\end{flushright}
\vspace{0.1in}

%\vspace{1cm}
\begin{center}
{\Large\bf\color{black}

Towards a  Gravity Dual  \\[0.5cm] for  the  Large Scale Structure 
of the Universe}\\
\bigskip\color{black}
\vspace{1cm}{
{\large A. ~Kehagias$^{a,b}$, A. ~Riotto$^c$ and M.~S.~Sloth$^d$}
\vspace{0.3cm}
} \\[7mm]
{\it {$^a$\, Physics Division, National Technical University of Athens\\ 15780 Zografou Campus, Athens, Greece}}\\
{\it {$^b$\, Theoretical Physics Department, CERN, CH-1211 Geneva 23, Switzerland}}\\
{\it $^c$ {Department of Theoretical Physics and Center for Astroparticle Physics (CAP)\\ 24 quai E. Ansermet, CH-1211 Geneva 4, Switzerland}}\\
{\it {$^d$\, CP$^3$-Origins, Center for Cosmology and Particle Physics Phenomenology\\
University of Southern Denmark, Campusvej 55, 5230 Odense M, Denmark}}\\
\end{center}
\bigskip

\vspace{.4cm}

\begin{abstract}
\noindent
The dynamics of the large-scale structure of the universe enjoys at all scales, even in the highly non-linear regime, a  Lifshitz symmetry during the matter-dominated period. In this paper we propose a general class of  six-dimensional spacetimes which could be a gravity dual to the four-dimensional large-scale structure of the universe. In this set-up,  the Lifshitz symmetry manifests itself as an isometry in the bulk and our universe is a four-dimensional brane  moving in such 
six-dimensional bulk. After finding the correspondence between the bulk and the  brane dynamical Lifshitz exponents,  we find the intriguing  result that the preferred value of the dynamical  Lifshitz exponent of our observed universe, at both linear and non-linear scales,
corresponds to a  fixed point of the   RGE flow  of the dynamical  Lifshitz exponent in   the dual system  where the symmetry is enhanced to the  Schr\"odinger group containing a non-relativistic conformal symmetry.
We also investigate  the RGE flow between fixed points of the Lifshitz dynamical exponent in the bulk and observe that this flow
is  reflected in a growth rate of the large-scale structure, which seems to be in qualitative agreement with what is observed in current data.  Our set-up might provide an interesting new arena for testing the ideas of holography and gravitational duals.

\end{abstract}
\bigskip

\end{titlepage}

\baselineskip=18pt
%%%%%%%%%%%%%%%%%%%%%%%%%%%%%%%%%%%%%%%%
\section{Introduction and summary \label{sec:intro}} 
%%%%%%%%%%%%%%%%%%%%%%%%%%%%%%%%%%%%%%%%
There is currently an intense experimental development of Large Scale Structure (LSS) surveys. The upcoming galaxy redshift surveys, such as Euclid \cite{euclid}, Dark Energy Spectroscopic Instrument (DESI) \cite{desi}, the Large Synoptic Survey Telescope (LSST) \cite{lsst}, and the Wide-Field InfraRed Survey Telescope (WFIRST) \cite{wfirst} are going to cover progressively larger fractions of the sky and deeper redshift ranges. 

The main method for generating theoretical predictions, which the data can be compared with, is traditionally large numerical N-body simulations. On the other hand, the analytical understanding of the density perturbations is poor. The main problem is that structure formation is an inherent non-linear problem, and at small scales perturbation theory completely breaks down due to the gravitational collapse. The reason is the following.
When the perturbations generated during inflation re-enter the horizon, they provide the seeds for the  LSS  of the universe and they  grow  via the gravitational instability \cite{Bernardeau:2001qr}. At early epochs, the growth of the density perturbations can be described by linear perturbation theory and the Fourier modes evolve independently from one another, thus conserving the statistical properties of the primordial perturbations. When the perturbations become non-linear, the coupling between the different Fourier modes become relevant, inducing nontrivial correlations that modify the statistical properties of the cosmological fields. At intermediate  scales the evolution of matter may be described analytically by extending the standard perturbation theory,  where one defines a series solution to the fluid
equations in powers of the initial density field. The $n$-th order term of the series for the density contrast grows as the $n$-th
power of the scale factor $a$ (for a pressureless fluid), thus  rapidly deteriorating  its convergence properties. 

The need for improving theoretical predictions for the next generation of very large galaxy
surveys has spurred many efforts to go beyond the standard perturbation theory. The renormalized perturbation theory \cite{rpt} reorganizes the perturbation expansion in terms of
different fundamental objects, the so-called non-linear propagator and non-linear vertices, to improve
the convergence. The renormalization group method \cite{rg1,rg2} represents an alternative possibility where truncating the renormalization equation  at the level of some $n$-point  correlator leads to a solution that corresponds to the summation of an infinite class of
perturbative corrections. 
The effective field theory of the LSS  \cite{Carrasco:2012cv}  is formulated  in terms of an IR effective fluid characterized by several parameters, such as speed of sound and viscosity and has attracted a lot of attention recently. Last, but not least, the  time-sliced perturbation theory makes use  of the  time-dependent probability distribution function to generate  correlators of the cosmological observables at a given moment of time \cite{Blas:2015qsi}.

It is fair to say that all the methods proposed so far break down on small scales in the truly non-linear regime and it seems that,  in order to analytically understand the LSS of the universe  on small scales, one  needs an intrinsically  non-perturbative approach, possibly based on symmetry arguments. 

It is well-known that weakly coupled dual descriptions of strongly-coupled conformal theories can be obtained through the  AdS/CFT duality \cite{mal,gub,wit}  expressed as 
gravity or string theory on a weakly curved spacetime. The symmetries of the gravitational
background  realize in a geometric manner  the symmetries of the dual field theory. 
The conformal group SO$(D, 2)$ of a $D$-dimensional CFT for instance  is connected to  the group of isometries of
${\rm AdS}_{D+1}$.

After the  original proposal of the AdS/CFT duality  \cite{mal}, many  phenomenological applications have been proposed,  for instance  the large viscosity of the quark-gluon plasma has been explained. More interestingly for us,   a strong-coupling description for non-relativistic systems has been proposed through a dual geometry whose isometries reproduce the symmetries of the non-relativistic conformal field theory. 
Gravity duals have been proposed for condensed matter systems whose symmetry is  the  non-relativistic conformal symmetry  \cite{Son:2008ye,Balasubramanian:2008dm,Adams:2008wt,kak}. These  systems are described by non-relativistic theories where the particle number can or cannot be conserved   and  are characterized by 
 a dynamical dynamical exponent $z\neq 1$. Systems enjoying this dynamical scaling are said to be Lifshitz symmetric. The dynamical exponent 
governs
the anisotropy between spatial and temporal scaling 
 
 \beq
t'= \lambda^z t,\,\,\,\, \vec{x}'=  \lambda \vec{x}.
\label{a}
\eeq
These considerations are  interesting   because the LSS  formation can be formulated as a non-relativistic field theory with the same Lifshitz symmetry
during the matter-dominated period and at any  scales.
It is natural to ask: can we find  a general class of higher-dimensional spacetimes which could
constitute a gravity dual to the four-dimensional LSS theory in such a way to match an otherwise strongly-coupled system into a weakly coupled gravity set-up?

In this paper we take the first step to answer this interesting and challenging question. Our  findings give us 
 reasons to conjecture that one  can understand the LSS  in the non-linear regime in terms of a weakly coupled gravitational system in six-dimensions with a metric of the form

\beq
\d s^2 =- \frac{L^{2 z}}{r^{2 z}}\d t^2+\frac{L^2}{r^2}\left(2\d t \d\xi+\d\vec{x}^2+\d r^2\right).
\eeq
 In this set-up, the Lifshitz symmetry manifests itself  as an isometry in the bulk and our universe is to be thought as  a four-dimensional brane immersed and moving  in a six-dimensional bulk. As we shall see, the six-dimensional  metric  realizes the Lifshitz symmetry of the four-dimensional LSS and is supported by a massive gauge field coupled to AdS gravity supplemented by a bulk scalar field playing the role of the holographic dual of dark matter.  Similarly to the case of the   AdS/CFT, the scale
in the dual field theory is mapped into  an extra radial dimension
on the gravity side of the duality, and the rescaling of this extra coordinate realizes 
the Lifshitz transformations (\ref{a}).

While the ultimate goal would be to find the dynamics of the relevant   boundary observables defined by the theory in the bulk  by
generalizing the usual holographic dictionary and to construct in this way boundary correlators (such as the one for the density contrast) as given by the value of the renormalized
bulk action for specified boundary values of the bulk fields, in this paper we will restrict ourselves to characterize the set-up. Our  preliminary  findings indicate that: 

\begin{itemize}

\item Without devising it, the dynamical Lifshitz exponent of our observed universe at  the linear and  non-linear scales turns out to match a    fixed point of the dynamical Lifshitz exponent of the dual system where the symmetry in the bulk is enhanced  to the  Schr\"odinger group containing a non-relativistic conformal symmetry. This is probably the main result of the paper. 
Despite the fact that this result  might be due to a pure coincidence, it is certainly an
 intriguing one. In the approximation of a pure matter-dominated universe, the so-called Einstein-de Sitter universe, the brane and bulk dynamical Lifshitz exponents do not evolve and the  full  Schr\"odinger symmetry in the gravity dual might represent a suitable starting point to investigate  the dark matter perturbations at all scales.

  \item  Inspired by the fact that non-relativistic four-dimensional theories with anisotropic scale invariance can flow under
relevant perturbation to Lifshitz-fixed points,   the gravitational dual of the LSS theory  comes with  Lifshitz fixed points and  the relevant
perturbations which  induce  Renormalization Group Evolution (RGE)   flows to these   Lifshitz fixed points. This flow  captures the small breaking of the Lifshitz symmetry during the
phase in which the universe is dominated by the vacuum energy till the dynamical exponent flows into the observed fixed point on very small scales. From the
four-dimensional point of view the relevant operator is given by the vacuum energy.

\item The holographic
RGE flow of the dynamical exponent  in the dual theory goes from a Lifshitz fixed-point  in the   UV  (at large $r$ but still $r\lsim L$)  to the same  value in  the  IR (at small  $r$).   In the four-dimensional side, this holographic flow corresponds to the evolution of the large-scale linear gravitational instabilities into the non-linear small scale LSS, which enjoys the Lifshitz symmetry.

 In Fig. 1 we schematically depict the set-up.
In a purely matter-dominated universe the fact that the brane  dynamical Lifshitz exponent remains constant is a consequence of a trivial
RGE flow in the bulk.

\item  The RGE flow of the dynamical exponent is reflected in a 
 growth of the LSS in the non-linear regime with roughly the same rate it has in  the linear regime and much slower than in the 
quasi non-linear regime. This result seems to correspond to what is observed in  current data.
\item The dark matter on the brane can  be realized in terms of dual  fields in the bulk, so that the non-relativistic system  on the brane inherits the symmetry properties of the bulk.
 \end{itemize}
\begin{figure}[h!]
    \begin{center}
      \includegraphics[scale=.6,angle=270]{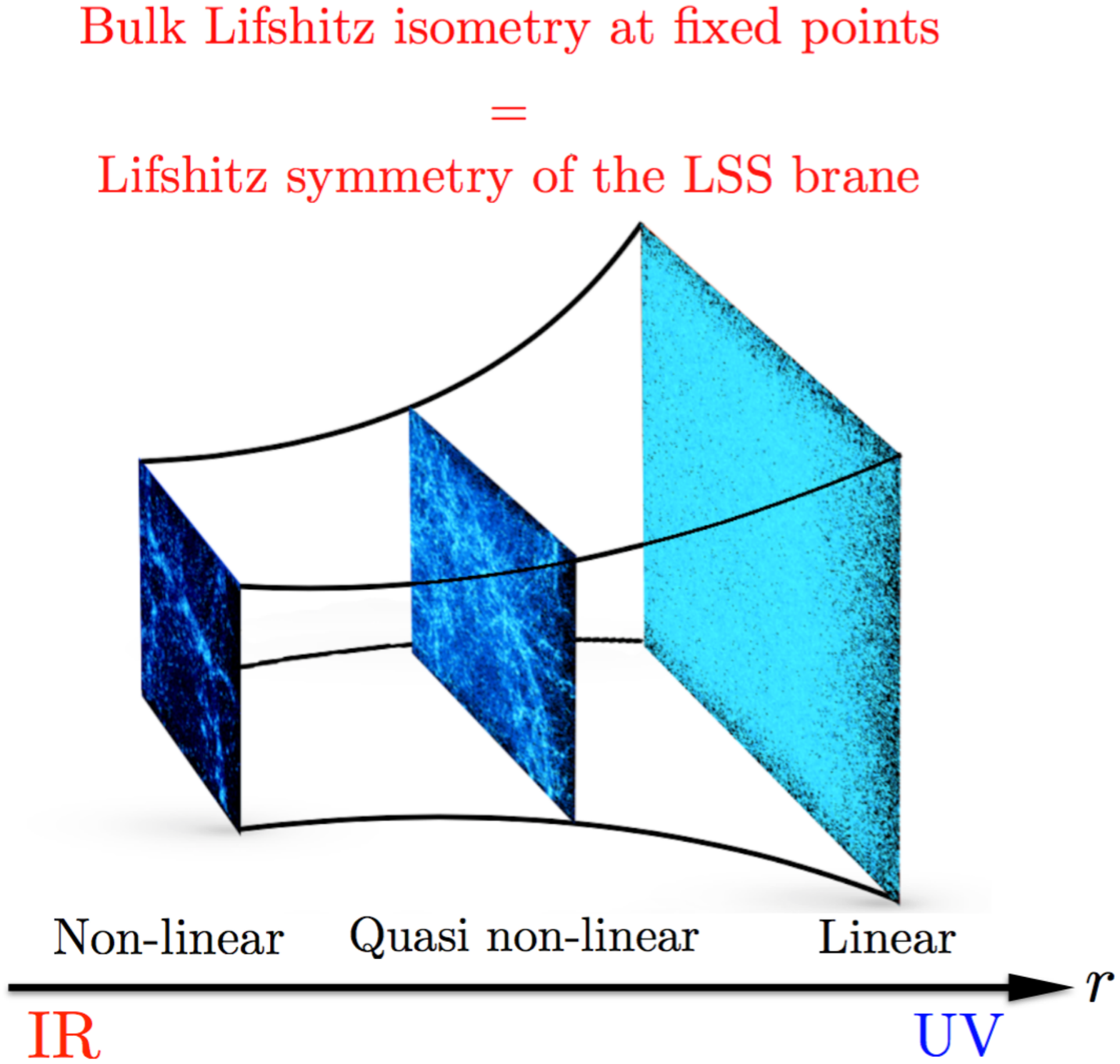}
    \end{center}
     \caption{\small A pictorial representation of our set-up. The six-dimensional bulk enjoys a Lifshitz isometry and realizes the Lifshitz symmetry of the dual field LSS theory on the our universe, the four-dimensional LSS brane, at the fixed points. The flow from large to small  $r$ of the bulk dynamical exponent $z$ describes the evolution from the large- to the small-scale LSS. The size of the slices transverse to $r$ are illustrated in comoving coordinates for a fixed physical size.}  
\end{figure}
 \noindent
The paper is organized as follows. In section 2 we describe why the Lifshitz symmetry is an exact symmetry, at any level of perturbation theory, for the dynamics of the dark matter during the matter-dominated period. In section 3 we identify the dynamical exponents of the Lifshitz symmetry for the LSS. In section 4 we describe the gravity dual. In section 5 we discuss the brane cosmology and  identify the correct matching between the values of the dynamical exponents at the bulk fixed points with the ones on the LSS brane in section 6. In section 7 we describe the Lifshitz RGE flow of the Lifshitz dynamical exponent and deal with the notion of holographic dark matter in section 8. We conclude in section 9. The paper is also supplemented by Appendix A where we   explain  how to identify the dual fields.
%and containing some comments about the special case $z=-3/2$ (appendix B).

\section{The LSS Lifshitz symmetry \label{sec:Boltzmann}}
In this section we show that the evolution of dark matter particles in a matter-dominated universe enjoys a Lifshitz symmetry. This is a standard result that goes back to Peebles \cite{peebles} and can be found in many good cosmology textbooks. We show it here for completeness and to remark that the Lifshitz symmetry is indeed valid at any scale.

Let us define the dark matter  (of mass $m$) phase space density $f(\vec{p},\vec{x},\t)$ in terms of the conformal time $\tau$, the comoving coordinates $\vec{x}$
and the canonical momentum  $\vec{p}= a m \vec{u}$ defined in terms of the peculiar velocity $\vec{u}$,  where $a(\t)$ is the scale factor of the expansion of the universe with rate ${\cal H}$.
The phase space describing the  self-gravitating gas of dark matter particles    obeys the following Boltzmann equation

\begin{eqnarray}
\frac{\del}{\del \t}f(\vec{p},\vec{x},\t)
+\frac{1}{am}\vec{p}\cdot\bnabla_{\vec{x}}  f(\vec{p},\bx,\t)-am\bnabla_{\vec{x}}\Phi\cdot\bnabla_{\vec{p}}f(\vec{p},\bx,\t)=0. \label{B1}
\end{eqnarray}
In addition, 
the canonical  momentum obeys the relation 
 \begin{eqnarray}
 \frac{{\rm d}\vec{p}}{{\rm d} \t}=-am\bnabla_{\vec{x}}\Phi(\vec{x},\t) \label{N}
 \end{eqnarray}
and the  Newtonian gravitational potential  $\Phi(\vec{x},\t)$ satisfies the Poisson equation
\begin{eqnarray}
\nabla^2\Phi(\bx,\t)=4\pi G\,  \overline \rho \,a^2
\delta(\bx,\tau),\,\,\,
\delta(\bx,\t)=\frac{\rho(\bx,\t)-\bar{\rho}}{\bar{\rho}},
 \label{Poisson}
\end{eqnarray}
where   $\bar{\rho}$ is the  the mean dark matter  density.
For a matter-dominated universe, the scale factor $a$ scales like $\tau^2$ and one can easily  verify that Eqs. (\ref{B1}), (\ref{N}) and (\ref{Poisson}) are invariant under the Lifshitz symmetry 

\beq
\boxed{\bx'=\lambda \bx, ~~~\tau'=\lambda^{\widetilde{z}}\t\label{as1}}\,,
\eeq
provided that the other quantities are transformed as follows

\begin{eqnarray}
f'(\vec{p},\vec{x},\t)&=&f(\vec{p}',\vec{x}',\t'),\nonumber\\ \label{f1p}
 \delta'(\bx,\t)&=&\delta(\bx',\t'),\label{ds1}\nonumber\\
 \vec{p}'(\bx,\t)&=&\lambda^{-(\widetilde{z}+1)}\vec{p}(\bx',\t'), \label{p1}\nonumber\\
 \Phi'(\bx,\t)&=&\lambda^{2(\widetilde{z}-1)}\Phi(\bx',\t'). \label{gfs1}
\end{eqnarray}
A few remarks are in order:

\begin{itemize}
\item The Lifshitz symmetry is valid only in a matter-dominated universe. When the cosmological constant (or quintessence) dominates and we have to deal with a two-component system, the symmetry is lost. Nevertheless,  the Lifshitz symmetry holds with a good approximation also at very small scales, where the dark matter density is much larger than the one provided by the vacuum energy. In this sense, the LSS at highly non-linear scales is Lifshitz symmetric. To get convinced about this point one can think about the   spherical collapse model where the scale factor is identified with
a local one describing the evolution of a matter-dominated universe with an effective positive curvature and the cosmological constant plays basically no role.  Also, matter  always largely dominates 
over the effective curvature term (in particular during the  the collapsing phase), only at the turning point the two contributions are equal.

\item As we already stressed, the Lifshitz symmetry is present at the level of Boltzmann equation. As such, it is valid on any scale and not only when the fluid approximation of dark matter holds. Even when we look at cosmological scales at which shell crossing and multiple streams
occur, the Lifshitz symmetry is there.

\item The Lifshitz symmetry is valid for any values of the Lifshitz exponent $\widetilde{z}$.

\item One should also keep in mind for the following that the Lifshitz dynamical exponent $\widetilde{z}$ in the four-dimensional system is not going to be identical to the bulk Lifshitz dynamical exponent $z$.  This is the reason why we adopt two different notations, $\widetilde{z}$ and $z$, to identify them.
\end{itemize}

\section{The LSS dynamical exponents \label{sec:critical}}
In this section we offer some considerations about the values of the Lifshitz dynamical exponents during the LSS evolution. The central quantity
in LSS cosmology is the power spectrum of the dark matter perturbations

\beq
P(k,\tau)=\big< \left|\delta_{\vec k}(\tau)\right|^2\big>.
\eeq
For convenience we will define also  the dimensionless power spectrum  

\beq
\Delta^2(k,\tau)=\frac{k^3}{2\pi^2}\,P(k).
\eeq
The evolution of the LSS starts with a linear power spectrum $\Delta_{\rm L}^2\lsim 1$ set by the primordial seeds generated during inflation and reprocessed once the perturbations re-enter the horizon. Gravitational instability drives the power spectrum to larger values in the so-called
Quasi Non-Linear (QNL) regime where $1\lsim \Delta_{\rm QNL}^2\lsim10^2$. During this phase perturbations grow faster than during the linear regime, till they enter in the so-called Non-Linear (NL) regime, $\Delta_{\rm NL}^2 \gsim 10^2$ when the growth of the perturbations slows down.

If  the Lifshitz  symmetry holds, the power spectrum must scale like 
\beq \label{Pform}
P(k,\tau)\equiv  \tau^{3/\widetilde{z}}  \mathcal{P}(k\tau^{1/\widetilde{z}}),
\eeq
where $\mathcal{P}$ is a priori an arbitrary (but possibly power-law)  function. In the following we will discuss what are the values of the critical Lifshitz exponent characterizing the three different phases. At this stage we do not aim at  a precise quantitative determination, but rather we aim at characterizing the qualitative behavior of the critical Lifshitz exponents.

\subsection{The linear regime}
In the linear regime the power spectrum is characterized by a spectral index $n$ and its growth is parametrized by the following expression

\beq
P_{\rm L}(k_{\rm L},\tau)\sim \, D^2(\tau) k_{\rm L}^n,
\eeq
where $D(\tau)$ is the linear growth factor.  In the matter-dominated period $D(\tau)=a(\tau)\sim \tau^2$ and 
by matching this behavior against the generic expression (\ref{Pform}), one finds

\beq
\label{as}
\boxed{\widetilde{z}_{\rm L}=\frac{n+3}{4}}\, .
\eeq
Of course, as soon as the vacuum energy  takes over, the Lifshitz symmetry is lost and therefore the value (\ref{as}) should be understood as the  initial condition (to be fixed in the matter-dominated period)
of the brane Lifshitz dynamical exponent $\widetilde{z}$ which, in turn, identifies the  initial condition
of the bulk  Lifshitz dynamical exponent $z$ in the dual theory. 

N-body simulations \cite{Lawrence:2009uk} indicate that the appropriate value  for our universe of $\widetilde{z}_{\rm L}$ when $\Delta_{\rm L}^2(k)={\cal O}(1)$ is obtained for 
$n\simeq -1.6$\footnote{We thank V. Desjacques for discussions about this point. }.  This leads to $\widetilde{z}_{\rm L}\simeq 1/3$. As we shall see,   this  value remarkably  corresponds in the gravity dual model    to a fixed point of the RGE flow  where the bulk symmetry is augmented by a  non-relativistic
conformal symmetry giving rise to the  Schr\"odinger group. 

\subsection{The quasi non-linear regime}
While the analytical methods fail in the QNL and NL  regimes, 
invaluable fitting  formulas calibrated  by simulations which describe how
the power spectra of density fluctuations evolve into the NL regime of
hierarchical clustering have been found in a series of remarkable papers \cite{ham,pd1,mo,pd2}. The foundation of these fitting formulas is the 
so-called HKLM  \cite{ham} relation which has
two parts. The first part is to connect the nonlinear wavenumber to the linear one

\beq
\label{lnl}
k_{\rm L}=\left[1+\Delta^2_{\rm NL}(k_{\rm NL},\tau)\right]^{1/3} \, k_{\rm NL}.
\eeq
The reasoning behind this is pair conservation.  The second part of the HKLM
procedure is to conjecture that the nonlinear correlation function is a universal function of
the linear one,

\beq
\Delta^2_{\rm NL}(k_{\rm NL},\tau)=f\left[\Delta^2_{\rm L}(k_{\rm L},\tau)\right].
\eeq
The asymptotics of the function $f(x)$ can be deduced readily. For small arguments $x\ll 1$,
$f(x)\simeq x$. Following the collapse of structures, $f(x)\simeq x^{3/2}$ derived from  the hypothesis that
although the separation of clusters will alter as the universe expands, their internal
density structure will stay constant with time.
Numerical experiments 
 interpolated between these two regimes, in a manner that empirically showed
negligible dependence on the power spectrum \cite{ham,pd1,mo,pd2}.

In the QNL  regime we can approximate the expression (\ref{lnl}) as

\beq
\label{lnls}
k_{\rm L}=\left[\Delta^2_{\rm QNL}(k_{\rm QNL},\tau)\right]^{1/3} \, k_{\rm QNL}.
\eeq
Furthermore, if we pose $f(x)=x^{1+\alpha}$, with $3.5\lsim \alpha \lsim 4.5$  \cite{pd1}, we have

\begin{eqnarray}
\Delta^2_{\rm QNL}(k_{\rm QNL},\tau)&\sim& D(\tau)^{(6-2\gamma_{\rm QNL})(1+\alpha)/3}\,k_{\rm QNL}^{\gamma_{\rm QNL}},\nonumber\\
\gamma_{\rm QNL}&=&\frac{3(3+n)(1+\alpha)}{3+(3+n)(1+\alpha)}.
\end{eqnarray}
Notice that in the QNL regime, the growth of perturbations speeds up. For $\gamma\simeq 1.5$\footnote{ Obtained when starting from the observed galaxy  index $\gamma_{\rm gal}\simeq 1.8$ for galaxies and rescaled taking into account the bias factor $b\simeq 1.2$ in $(\Delta^2)^b=\Delta^2_{\rm gal}\sim k^\gamma_{\rm gal}$, with $b\simeq 1.2$ \cite{pbook}.}  and $\alpha\simeq 4$, the perturbations scale like $D^5$, much faster than what they do on the linear regime. 

 If we  phenomenologically parametrize the  linear growth factor as $D(\tau)\sim \tau^{2+\epsilon}$ with a constant $\epsilon>0$  when the vacuum energy dominates, then one can define the QNL dynamical exponent

\beq
\widetilde{z}_{\rm QNL}=\frac{n+3}{4(1+\epsilon)}.
\eeq
This is a fake dynamical exponent  as the Lifshitz symmetry is slightly broken, but the interesting point is that it decreases with respect to the value during the linear phase. We will argue  that in the 
dual theory this corresponds to the running of the  bulk Lifshitz dynamical exponent to  values larger than the fixed point.

\subsection{The  non-linear regime}
In the NL regime, even when the vacuum energy dominates the effective expansion at small scales is governed by dark matter. Repeating the computations of the previous subsection with $\alpha=1/2$ and assuming that the  Lifshitz symmetry holds, one finds 

\begin{eqnarray}
\Delta^2_{\rm NL}(k_{\rm NL},\tau)&\sim& \tau^{(6-2\gamma_{\rm NL})}\,k_{\rm NL}^{\gamma_{\rm NL}},\nonumber\\
\gamma_{\rm NL}&=&\frac{3(3+n)}{5+n}.
\end{eqnarray}
This corresponds to the  value of the dynamical exponent 

\beq
\label{ass}
\boxed{
\widetilde{z}_{\rm NL}=\frac{n+3}{4}}\, .
\eeq
The fact that this value matches the linear one in Eq. (\ref{as}) is of course obvious because the Boltzmann equation describing the
evolution of the dark matter particles at non-linear scales experiences a matter-dominated environment. The value (\ref{ass})
 should be understood as the  final condition
of the brane Lifshitz dynamical exponent $\widetilde{z}$ which, in turn, identifies the  final  condition
of the bulk  Lifshitz dynamical exponent $z$ in the dual theory. 

A satisfactory gravity dual will then possess the following property: if the  RGE flow  when going from the UV to the IR in the bulk causes the 
  the bulk Lifshitz dynamical exponent to run from a fixed point back to the same fixed point by acquiring different  values during the intermediate evolution, this should be reflected in a change
of the LSS  Lifshitz dynamical exponent from its value at linear scales back to the same value at highly non-linear scales by  decreasing its value
in the intermediate evolution. We will show in the following that such a gravity dual  can indeed be constructed.

\section{The LSS gravity dual \label{sec:dual}}
Having identified the critical Lifshitz exponents for the LSS, we now proceed to construct the 
gravity dual theory possessing the appropriate Lifshitz symmetry. As we already mentioned in the introduction, the following metric

\beq
\label{m01}
\d s^2 =- \frac{L^{2 z}}{r^{2 z}}\d t^2+\frac{L^2}{r^2}\left(2\d t \d\xi+\d\vec{x}^2+\d r^2\right)
\eeq
is Lifshitz symmetric. In particular, it is symmetric under the scaling\footnote{Let us also notice that the speed of light $c$ in this set-up is $r$-dependent. Indeed, at constant $(\xi,x_2,x_3)$, one gets
$c=\left(L/r\right)^{z-1}$. 
For $z>1$,  we have $c\to \infty$ as $r\to 0$, a signal of non-relativistic theory at the boundary.  
%On the other hand, for $z<-1$, we find that $c\to \infty $ in the opposite 
 %$r\to \infty$ limit. In particular,   $c=0$ at $r=0$: the boundary at $r=0$ is
  %like the horizon of   a black hole. 
%Similarly to the latter,  the $r=0$ boundary is also an infinite redshift
 %surface.
}

\begin{eqnarray}
t\to \lambda^z t, ~~~r\to \lambda r, ~~~\vec{x}\to \lambda \vec{x}, ~~~\xi\to \lambda^{2-z}\xi.  \label{scal}
\end{eqnarray}
In fact, the metric (\ref{m01}) has more symmtries. It is invariant under an eleven-dimensional group of isometries consisting of translations in $\vec{x}$ and  $t$, rotations of
$\vec{x}$, Lifshitz transformations,  and  Galilean boosts 

\beq
t\to t'=t, \,\, \vec{x}\to\vec{x}'=\vec{x}-\vec{v} t,\, \,  
\xi\to \xi'=\xi+\frac{1}{2}\left(2\vec{v}\cdot \vec{x}-v^2t\right).
\eeq
What is important to notice is  that  $z=2$ is a point of enhanced symmetry. Indeed for such a value,  there is an additional   special conformal isometry

\beq
t\to t'=\frac{t}{1+\alpha t},\,\,  \vec{x}\to\vec{x}'=\frac{\vec{x}}{1+\alpha t},\,\, r\to r'=\frac{r}{1+\alpha t},\,\,
\xi\to \xi'=\xi+\frac{\alpha}{2}\frac{x^2+r^2}{1+\alpha t}.
\eeq
This special conformal symmetry  together with the symmetries above form the Schr\"odinger group \cite{Nishida:2007pj}.  This point will be come relevant in the following as we shall show that the LSS dynamical exponent for our observed universe at linear and non-linear scales corresponds to a bulk dynamical exponent $z$ intriguingly close to the point of enhanced symmetry.

From the metric (\ref{m01}) one can identify what  the running of $r$ corresponds to in terms of the physical length scales on a brane located at a given $r$: being $g_{ij}=(L^2/r^2)\delta_{ij}$,  we read that the ``redshift" factor scales like $1/r$. This implies that 
 that a brane observer at some reference scale $r_*$ will determine a physical length as  $\ell_*=(L_*/r_*) \ell_{\rm com}$. At another point $r=R(t)$, the same mode will look having a different $\ell=(L/R)  \ell_{\rm com}$. Therefore, we get that 
 
 \beq
 \label{sdf}
 \boxed{
 \ell=\frac{r_*}{R(t)} \ell_*}\, .
 \eeq
 If our universe is then identified with a moving brane with posotion $R(t)$  in the bulk, going from large to small values of $r$ corresponds to go from UV to the IR part in the bulk. In section 5  we will show that $R(t)$ will correspond with the inverse of the induced scale factor on the brane when the universe is matter-dominated. 
%  In particular, we are interested in those LSS modes for which the motion of the brane in the bulk takes a wavenumber in the linear regime during the matter dominated period and transports it into the QNL regime during the phase when the universe is dominated by the vacuum energy, down to
% the very NL regime where locally the universe is dominated the dark matter.
% 
 
 The metric  (\ref{m01}) is the solution of a massive Maxwell field coupled to AdS gravity
 \cite{Son:2008ye,Balasubramanian:2008dm}. Indeed, let us consider  the action
\beq
S= \int \d^6 x  \sqrt{-g}\left(R-2\Lambda-\frac{1}{4}F_{\mu\nu}F^{\mu\nu}+\frac{1}{2}m^2A_\mu A^\mu\right).
\eeq
The  equation for the massive vector $A_\mu$ is 
\beq
\p_{\nu}\left(\sqrt{-g}g^{\nu\rho}g^{\mu\sigma}F_{\rho\sigma}\right)-m^2\sqrt{-g}g^{\mu\nu}A_\nu=0~. \label{proca}
\eeq
%We can think of the mass term as a current $j_\mu = m^2 A_\mu$, and if we think of constant density $j_0=m^2 A_0$, with the ansatz $A_0=\propto r^{-z}$, then the Proca equation takes the form
%\beq
%\p_r\left(\sqrt{-g} g^{rr} g^{\xi 0}F_{r0}\right) -m^2\sqrt{-g}g^{\xi0}A_0=0
%\eeq
%and using $\sqrt{-g}= L^6/r^6$ implies that $A_0 \propto r^{-z}$ is a solution if
Symmetry considerations lead one to look for solutions for the gauge field of the type 
\begin{eqnarray}
A_\mu =A_0(r) \delta^0_\mu.
\end{eqnarray}
The field Eq. (\ref{proca}) imposes  
\begin{eqnarray}
A_0 =\frac{\rho_0}{z(z+3)} \frac{1}{ r^z}, \label{a0}
\end{eqnarray}
 where $\rho_0$ is  constant if we choose
\beq
m^2 = \frac{1}{L^2} \left[z(z+3)\right]. \label{mz}
\eeq
Let us note that the above relation is invariant under 

\beq
z\to -z-3.
\eeq  
The constant $\rho_0$ can be specified by Einstein equations. It turns out that for the metric (\ref{m01}) we have  the following components of the Einstein tensor
%Now, from the $0-0$ component of the Einstein equation, we can determine the constant of %proportionality and completely fix $A_0$. If we write $A_0= \rho_0/(z(z+3)) r^{-z}$, then the  %$0-0$ component of the Einstein equation will determine $\rho_0$. The Einstein tensor can b%e computed to be
\beq
G_{00} = (-13+z+2z^2)\frac{1}{L^2}\frac{L^{2z}}{r^{2z}},\qquad G_{0\xi}=G_{xx}=G_{yy}=G_{zz}=G_{rr}=\frac{10}{r^2}.
\eeq
This implies that the energy-momentum tensor must take the form \cite{Son:2008ye,Balasubramanian:2008dm}
\beq \label{T1}
T_{\mu\nu} = -\Lambda g_{\mu\nu} -{\mathcal{E}}\delta_\mu^0\delta_\nu^0g_{00},
\eeq
with 
\begin{eqnarray}
\Lambda =-\frac{1}{L^2}, ~~~{\mathcal{E}} =\frac{2 z^2 +z-3}{L^2}.
\end{eqnarray}
 We can compare this with the energy-momentum tensor of the vector field
\beq
T^{\mu\nu}= F^{\mu\be}F^\nu_\be -\frac{1}{4}g^{\mu\nu}F_{\al\be}F^{\al\be}-m^2 A^\mu A^\nu +\frac{1}{2}g^{\mu\nu} m^2 A_\rho A^\rho~,
\eeq
%whose $0-0$ component is
%\beq
%T_{00} = \frac{1}{2} g^{rr} \p_r A_0 \p_r A_0 +\frac{1}{2} m^2 A_0^2~.
%\eeq
%inserting 
%\begin{eqnarray}
%A_0= \rho_0/(z(z+3)) r^{-z} \label{a0}
%\end{eqnarray}
which, for the field in Eq. (\ref{a0}) matches  (\ref{T1}) if the constant $\rho_0$ is 

\beq
\rho_0^2={\mathcal{E}}\frac{2z(z+3)^2}{2 z+3}L^{2z+2}=2z(z-1)(3+z)^2L^{2z}.
\eeq
%Note that we should have 
%\begin{eqnarray}
%z\geq 1,
%\end{eqnarray}
%in order to have a real field $A_\mu$. 
%
Having constructed the appropriate gravity dual enjoying the Lifshitz symmetry (\ref{a}), 
we now study the four-dimensional brane cosmological and matching between the brane and bulk Lifshitz dynamical exponents. We will then 
 proceed to show that
our gravitational dual of Lifshitz-like theories come with the  relevant
perturbations that induce the  flow of the dynamical exponent towards a fixed point. 
 This flow is reflected in the evolution of the brane Lifshitz dynamical exponent from the initial condition  $\widetilde{z}_{\rm L}$ for the linear perturbations down to 
the final fixed point value  $\widetilde{z}_{\rm NL}$ on small scales passing through $\widetilde{z}_{\rm QNL}$ under the action of the relevant operator represented by the four-dimensional cosmological constant. In other terms, the bulk flow  accounts for  
 the small breaking of the Lifshitz symmetry during the
phase in which the universe is dominated by the vacuum energy till the dynamical exponent flows into a fixed point on very small scales.

 \section{Brane Cosmology}
 Let us now consider a four-brane floating around in the background (\ref{m01}) whose motion will induce a cosmological evolution on the branes \cite{Kr,KK}. According to the standard nomenclature, a four-brane is a five-dimensional hypersurface embedded here in the six-dimensional background.   The coordinates of the four-brane are then $(\zeta^0,\zeta^i,\zeta^4) ~(i=1,2,3)$ and let us 
  assume that its  position  in the gauge $(\zeta^0=t,\zeta^i=x^i,\zeta^4=\xi)$ is at $r=R(t)$. In other words, the four-brane moves perpendicular  to the $r$-direction. In this case,
for a generic metric of the form
\begin{eqnarray}
 \d s^2=-f(r) \d t^2+g(r)\left(2\d t \d \xi+\d\vec{x}^2\right)+h(r) \d r^2,  \label{mee}
 \end{eqnarray} 
  the brane velocity is $u^\mu$ (normalized to $u^\mu u_\mu=-1$)   is 
  given by 
 \begin{eqnarray}
 u^t=\left(\frac{h\dot{R}^2}{f}+\frac{1}{f}\right)^{1/2}, ~~~u^r=\dot{R}. 
 \end{eqnarray}
The induced metric on the brane is 
 \begin{eqnarray}
 \d s^2_{\rm ind}=-\d \eta^2+g(R(\eta))\left(2\frac{\d \eta}{f^{1/2}(R(\eta))}\d \xi+\d\vec{x}^2\right),  \label{gind}
 \end{eqnarray}
and  $\dot{R}=\d R/\d\eta$. 
 The normal to the brane satisfies the equation $n^\mu u_\mu=0$ and it is turns out to be 
 \begin{eqnarray}
 n^t=-\sqrt{\frac{h}{f}} u^r,~~~n^r=-\sqrt{\frac{f}{h}}u^t.
 \end{eqnarray}
 The extrinsic curvature is given by 
 \begin{eqnarray}
 K_{\mu\nu}=\nabla_\mu n_\nu
 \end{eqnarray}
 and the Israel matching condition is 
 \begin{eqnarray}
  K^+_{\mu\nu}-K^-_{\mu\nu}=-\frac{1}{M_*^{4}} \left(T_{\mu\nu}-\frac{1}{4} T^\lambda_\lambda \gamma_{\mu\nu}\right),   \label{israel}
  \end{eqnarray} 
 where $K^\pm_{\mu\nu}$ is the extrinsic curvature left and right of the brane, $T_{\mu\nu}$ is the energy-momentum tensor on the brane, $\gamma_{\mu\nu}$ is the induced metric and $M_*$ is the six-dimensional Planck mass. 
 %For a brane with only tension $T_{\mu\nu}=-\sigma \gamma_{\mu\nu}$ we find (I think!) for the %Lifshitz background (\ref{m01}) that 
 %\begin{eqnarray}
% \end{eqnarray}
 The extrinsic curvature for the metric (\ref{mee}) is found to be 
 \begin{eqnarray}
 K^\pm_{\mu\nu}=\mp \left(\frac{1}{h_\pm}+\dot{R}^2\right)^{1/2}\frac{g'}{2g}. 
 \end{eqnarray}
 In addition,  a four-brane with tension $\sigma$ and  a perfect fluid on its five-dimensional world-volume with energy density $\rho_5$ and pressure $P_5$  
 the energy-momentum tensor is
 \begin{eqnarray}
  T_{\mu\nu}=-\sigma \gamma_{\mu\nu}+\rho_5 U_\mu U_\nu +P_5\left(\gamma_{\mu\nu}+U_\mu U_\nu\right). 
  \end{eqnarray} 
Hence,   Eq. (\ref{israel}) is explicitly written as 
  \begin{eqnarray}
  \left(\frac{1}{h_+}+\dot{R}^2\right)^{1/2}+\left(\frac{1}{h_-}+\dot{R}^2\right)^{1/2}=
  \frac{1}{4M_*^{4}}(\sigma+\rho_5) \frac{2g}{g'}.  \label{is}
  \end{eqnarray}
 %From (\ref{is}) we find that 
 %\begin{eqnarray}
 %\dot{R}^2=\frac{1}{64M_*^8 (\rho+\sigma)^2}\bigg\{
 %256 f_+^2+\Big{(}-16 f_-+M_*^{-8}8(\rho+\sigma)^2\Big{)}^2-32 f_+\Big{(}16 f_- %+M_*^8(\rho+\sigma)^2\Big{)}\bigg\}
% \end{eqnarray}
 %where 
 %\begin{eqnarray}
 %ds^2_\pm =-f_\pm(r) dt^2+g_\pm(r)\left(2dtd\xi+d\vec{x}^2\right)+h_\pm(r) dr^2 \label{mee1}
 %\end{eqnarray} 
 For a Lifshitz background 
 \begin{eqnarray}
 f_\pm=\frac{L^{2z_\pm}}{r^{2z_\pm}}\, , ~~~~g_\pm=h_\pm=\frac{L^2}{r^2}.
 \end{eqnarray}
 In the particular case $z_+=z_-$, we find that 
 \begin{eqnarray}
 \dot{R}^2=-\frac{R^2}{L^2}+\frac{R^2}{64 M_*^{8}} (\rho_5+\sigma)^2. \label{frw}
 \end{eqnarray}
 We may write this equation as a Friedmann equation after we identify the scale factor and recall  that we need to perform  a further Kaluza-Klein along the $\xi$-direction\footnote{ 
The latter is a null direction and compactification is quite involved in this case \cite{NJ} and a common problem for all the  condensed matter systems described through gravity duals. The main problem here is  the zero mode of the momentum along the null direction (identified with the particle number) as  discussed in \cite{Pol}. This leads to certain issues with the metric
  (\ref{m01}) with a compact null direction since there is a vanishing spatial circle in this case  \cite{MMT}. These issues are also related  to the discreteness of  the associated particle number. Here we will simply  assume that there is an effective four-dimensional  description of the dynamics,  {\it i.e.}  that ${\cal L}_{\rm eff} ^{(4)}$ in Eq. (\ref{4def}) is finite. }.
 As it is usual in the Kaluza-Klein set-up, the induced metric
in Eq. (\ref{gind}) 
 is not in the appropriate Einstein frame and some redefinition is needed. 
  %As we will need to dimensional reduce along this null direction, the five dimensional brane %action will be of the form
The five dimensional action is 
\begin{eqnarray}
{\cal S}= \int {\rm d} ^3 x{\rm d \xi}{\rm d}\eta\, \sqrt{-g_{\rm ind}}\,{\cal L}_{\rm eff}^{(5)}, 
\end{eqnarray}
where 
\begin{eqnarray}
\sqrt{-g_{\rm ind}}=\left(\frac{R}{L}\right)^{z-5}.
\end{eqnarray}
Defining a four-dimensional effective action
\begin{eqnarray}
{\cal L}_{\rm eff}^{(4)}=\int {\rm d}\xi\, {\cal L}_{\rm eff}^{(5)},  \label{4def}
\end{eqnarray}
 we get that 
\begin{eqnarray}
{\cal S}= \int {\rm d} ^3 x{\rm d}\eta \left(\frac{R}{L}\right)^{z-5}\, {\cal L}_{\rm eff}^{(4)}=\int {\rm d} ^3 x{\rm d}\eta\, \sqrt{-g_{\rm E}}\, {\cal L}_{\rm eff}^{(4)}. 
\end{eqnarray}
Therefore, the four-dimensional action can be reproduced through 
a four-dimensional  Einstein metric 
\begin{eqnarray}
\d s^2_{\rm E}=-\left(\frac{R}{L}\right)^{2z-4}{\rm d}\eta^2+\left(\frac{L}{R}\right)^{2}
{\rm d}\vec{x}^2.  \label{ein00}
\end{eqnarray}
We can write Eq. (\ref{ein00}) as 
\begin{eqnarray}
\d s^2_{\rm E}=\left(\frac{L}{R}\right)^{2}\bigg[-\left(\frac{R}{L}\right)^{2z-2}{\rm d}\eta^2+
{\rm d}\vec{x}^2\bigg]=a^2(\tau)\left(-\d\tau^2+{\rm d}\vec{x}^2\right), 
\end{eqnarray}
where we have defined  the four-dimensional scale factor 
\begin{eqnarray}
\label{df}
a=\frac{L}{R}, 
\end{eqnarray}
and  the conformal time through the relation  
\begin{eqnarray}
\label{etatau}
\d\tau=a^{1-z} \d\eta. 
\end{eqnarray}
Eq. (\ref{frw}) takes the form 
 \begin{eqnarray}\label{Hb}
 \left(\frac{\dot{a}}{a}\right)^2=-\frac{1}{L^2}+\frac{1}{64 M_*^{8}} (\rho_5+\sigma)^2.  \label{fried}
 \end{eqnarray}
 For the particular case of 
 \begin{eqnarray}
 \sigma=\frac{8M_*^4}{L},
 \end{eqnarray}
 we get the Friedmann equation five-dimensional (expressed in terms of the four-dimensional scale factor)
 \begin{eqnarray}
 \left(\frac{\dot{a}}{a}\right)^2=\frac{1}{4M_*^{4}L}\, \rho_5+\frac{1}{64 M_*^{8}} \rho_5^2\, . \label{frd}
 \end{eqnarray}
Note that $\rho_5$ is the five-dimensional energy density. We may assume that the extra fifth coordinate is a KK compact direction with  radius $R_5$. In that case, the zero-mode reduced four-dimesional energy density $\rho$ will be given by 
\begin{eqnarray}
\rho=2\pi R_5\rho_5.
\end{eqnarray}
 Therefore, to leading order in $1/R_5^2$ we will have 
 \begin{eqnarray}
 \left(\frac{\dot{a}}{a}\right)^2=\frac{8\pi G_{\rm N}}{3} \rho, \label{fried1}
 \end{eqnarray}
where 
\begin{eqnarray}
8\pi G_N=\frac{3}{12\pi L R_5 M_*^{4}}. 
\end{eqnarray}
Therefore, the standard  FRW dynamics is recovered on the brane for the four-dimensional zero-modes of the  five-dimensional KK brane at $r=R(\tau)$. 
At this stage, let us remark that the four-dimensional brane cosmology has been obtained by adding  an extra-dimensional
source of energy $\rho_5$ with the appropriate scaling of the scale factor to achieve the correct four-dimensional  universe evolution. While this is the standard procedure in the literature and is perfectly legitimate, it would be desirable to realize the dark matter on the brane by slightly altering the bulk geometry  in such a way that the dark matter  on the brane  inherits the symmetry properties of the bulk. We will come back to this point later on.

\section{The correspondance between the bulk and the brane  Lifshitz dynamical exponents}
The next step is to  find the correspondance between  the brane dynamical exponent $\widetilde{z}$ and the  bulk dynamical exponent $z$. To do so, we  express the coordinate time $\eta$, which  is invariant under Lifshitz scaling,  in terms of the conformal time $\tau$, which scales as in Eq. (\ref{as1}). 
%The definition of the  four-dimensional  conformal time is not straightforward as it involves the  fifth null direction $\xi$, which is accompanying by a radion field proportional to $R^{z-1}$. 
First of all, going back to Eq. (\ref{sdf}), we see that the running of $r$, identified with the movement of the LSS brane in the bulk, corresponds to 
a change of the physical length according to the relation

 \beq
 \label{sdfq}
 \ell(t)=\frac{a(t)}{a_*} \ell_*.
 \eeq
%The brane metric turns out to be
%\begin{eqnarray}
%\d s_{\rm E}^2=-\left(\frac{R}{L}\right)^{4-2z} \d t^2+\frac{R^2}{L^2} \d\vec{x}^2.
%\end{eqnarray}
%Then, we find that the coordinate time $t$ is related to conformal time $\tau$ by 
%\begin{eqnarray}
%\d t=a^{z-1} \d\tau.
%\end{eqnarray}
For the  matter-dominated period which is of interest for us, $a(\tau)\sim \tau^2$ and using Eq. (\ref{etatau}) we get 

\beq
\eta\sim \tau^{2z-1}.
\eeq
Therefore, recalling that $\eta$ is invariant  and that $\tau$ transforms into $\lambda^{\widetilde{z}}\tau$, we find
\begin{eqnarray}
\widetilde{z}=\frac{1}{2z-1}.
\end{eqnarray}
As we have previously  noticed, the point $z=2$ is a point of enhanced symmetry.  At this point, the value of $\widetilde{z}$  turns out to be $\widetilde{z}=1/3$

\beq
\boxed{{\rm Bulk}\,\,{\rm enhanced}\,\,{\rm symmetry}\,\,{\rm point}\,\, z=2\,\,\Rightarrow \,\,\widetilde{z}=\frac{1}{3}\,\, {\rm on}\,\,{\rm the}\,\,{\rm brane}\,\, \Rightarrow \,\, n=-1.6}\, .
\eeq
This is one of the main results of the paper and quite exciting because
this value of the  LSS dynamical exponent  $\widetilde{z}=(n+3)/4$ is obtained exactly for the spectral index $n=-1.6$,  
which is approximately  the value of the spectral index  for our observed universe when $\Delta^2_{\rm L}$ becomes of order unity. 

In the approximation of a pure matter-dominated universe, the so-called Einstein-de Sitter universe, the dynamical Lifshitz exponents, both on the brane and in the bulk, are constant and the  description of the LSS through a gravity dual with the full Schr\"odinger symmetry might represent a suitable starting point to characterize the dark matter perturbations.

%-------BRANE TIME--------
%
%
%We are interested in the relation between the induced time on the brane, $\eta$, and the coordinate time in the bulk metric $t$. From the identification
%\beq
%-d\eta^2 = -f(R(t))dt^2 +h(R(t))dR^2(t) 
%\eeq
%we have 
%\beq
%\frac{dt^2}{d\eta^2} = \frac{1}{f(R)}+\frac{h(R)}{f(R)}\dot R^2~.
%\eeq
%Using $a=L/R$, $H=\dot a/a =-\dot R/R$, $f =a^{2z}$, and $h=a^2$, we have
%\beq
%dt = \sqrt{a^{-2z}(1+H^2)}d\eta~,
%\eeq
%so for $H<<1$, we do have the simpler relation
%\beq
%dt = a^{-z}d\eta
%\eeq
%and defining a comoving brane time $ad\tau = d\eta$ lead to $dt =a^{1-z}d\tau$ -- which is similar but not quite the same as (4.32) above.
%
%Also, not here $\eta$ is the five dimensional brane time. How it relates to the actual physical time of Large Scale Structure further depends on the dimensional reduction on the brane from 5 to 4 dim. Although if we use that $\xi$ is compact and $\rho$ is the zero KK mode of $\rho_5$, this part might not change the relationship. However, if we use the older elegant example, where the 4-D DM fluid comes from dimensional reduction of a massless scalar field in 5-D, then we need to get of the $b$ multiplying the $d\eta d\xi$ term in (4.25). This can f.ex. be done by taking $\xi \to \xi/b$, but this will also shift the time wit an additional factor $(1+\dot b/b^2)$. If $b=2a^{2-z}$ (for $H<<1$) then it depends on the sign of $z$ if this term can be ignored.
%
\section{Lifshitz flows of the dynamical exponent \label{sec:running}}
Having constructed the gravity dual possessing the Lifshitz isometry and having identified the mapping between the Lifshitz bulk dynamical exponent
with the one on the four-dimensional brane, we now
study the renormalization group flow between   fixed points. This flow is supposed to capture the departure from the
exact Lifshitz symmetry which occurs when the universe is not matter-dominated. 
In the dual theory the variable $r$ goes from an  UV   initial condition at large $r$,  identified with the   Lifshitz fixed-point when  the linear perturbations are in the matter-dominated period,   to an IR  Lifshitz fixed-point at small $r$. 

A simple Langrangian   which is sufficiently general to capture the flow is the one where the latter is  triggered and controlled by a scalar field $\phi$ with potential $V(\phi)$ and coupling $W(\phi)$ to the gauge field

%In order to achieve the running of the bulk dynamical exponent, This means that we should have a dynamical system with $z=z(r)$, which will have fixed points $z_*$. Then the system will flow from one fixed point to another and depending on the nature of latter (if it is attractive or repulsive) will stay there or will return to the original one for example. We will discussed the simplest case where the running is triggered and controlled by a scalar field $\phi$ with potential $V(\phi)$. Moreover, we will assume that the scalar is coupled to the gauge fields through a mass function $W(\phi)$ so that the action we will consider here is of the form 
%Let us now consider a  vector and a scalar coupled to gravity with action
\begin{eqnarray}
 {\cal S}=\int \d^6 x\sqrt{-g}\bigg(R-2 V(\phi)-\frac{1 }{2}(\partial\phi)^2 -\frac{1}{4}
 F_{\mu\nu}F^{\mu\nu}-\frac{1}{2}W(\phi) A_\mu A^\mu\bigg).\label{s1}
 \end{eqnarray} 
%Note that we have allowed for instabilities by introducing $s_{\phi,A}=\pm 1$.  
A metric ansatz which is sufficiently general to capture this flow is
\begin{eqnarray}
  \d s^2=-e^{2A(y)}\d t^2+e^{2B(y)} \Big{(}2 \d t \d\xi+\d\vec{x}^2\Big{)}+\d y^2 \label{RSf}
  \end{eqnarray}  
  and in addition
  \begin{eqnarray}
  \phi=\phi(y), ~~~A_\mu=H(y) e^{A(y)}\delta_\mu^0.
  \end{eqnarray}
The coordinate $y$ above is related to $r$ defined in Eq. (\ref{m01}) as 

\beq
\label{cvb}
 \d r^2=e^{-2B(y)} \d y^2. 
\ee
The equations of motion can be expressed in terms of the scalars $\phi(y)$, $H(y)$ and 
\begin{eqnarray}
L(y)=\frac{1}{B'(y)}, ~~~z(y)=\frac{A'(y)}{B'(y)}, \label{LB}
\end{eqnarray}
(where prime denotes derivative with respect to $y$) as 
\begin{eqnarray}
0&=& \frac{z'L-L'z}{L^2}H+\frac{z^2}{L^2}H+\frac{2z}{L}H'+\frac{3}{L}\Big{(}H'+\frac{z}{L}H\Big{)}+H''-W H, \label{eq1}\\
0&=&\phi''+\frac{5}{L}\phi'-2\partial_\phi V ,\label{eq2}\\
0&=& z'L+(1-z)L'+(3+2z)(z-1)-\frac{1}{2}L^2\Big{[}\left(H'+\frac{z}{L}H\right)^2+ W H^2\Big{]},\label{eq3}\\
0&=&\frac{1}{2}\phi'^2-\frac{4}{L^2}L', \label{eq4}\\
0&=&\frac{1}{4}\phi'^2-V-\frac{10}{L^2}. \label{eq5}
\end{eqnarray} 
It is easy to check that 
for the particular case 
\begin{eqnarray}
 W=\frac{z_0(z_0+3)}{L_0^2}, ~~~V=-\frac{10}{L_0^2}, ~~~\partial_\phi V(\phi_0)=0,~~~L=L_0, \label{WV0}
 \end{eqnarray} 
 the solution is 
 \begin{eqnarray}
 A=\frac{z_0 y}{L_0}, ~~~B=\frac{y}{L_0}, ~~~\phi=\phi_0, ~~~
 H=\sqrt{\frac{2(z_0-1)}{z_0}},
 \end{eqnarray}
which is just Eq. (\ref{m01}).  
%This is a very particular solution and it is only valid for the constant potential $V$  in (\ref{WV0}). For any other  constant value of $V$ and %constant $W$, 
% $\phi,L$ will have a not trivial profile, as can be seen from Eq. (\ref{eq5}) and then, both $L$ and $z$ will be running. 
 The system of Eqs.  from (\ref{eq1})  to (\ref{eq5}) has  four fixed points at \cite{Liu:2012wf,Liu:2015xxa}
\begin{eqnarray}
H&=&0,\,\, z=-\frac{3}{2},\nonumber\\
H&=&0,\,\,z=1,\nonumber\\
H^2&=&\frac{1}{L^2W}\bigg(2L^2 W-3- \sqrt{9+4 L^2 W}\bigg),\,\, 
z=-\frac{3}{2}\bigg(1-\sqrt{1+\frac{4L^2 W}{9}}\bigg),\nonumber\\
H^2&=&\frac{1}{L^2W}\bigg(2L^2 W-3+ \sqrt{9+4 L^2 W}\bigg),\,\, 
z=-\frac{3}{2}\bigg(1+\sqrt{1+\frac{4L^2 W}{9}}\bigg).
\label{fixedpoints}
\end{eqnarray}
It can easily be checked that for the forth  fixed point
\begin{eqnarray}
 W\geq 0,  ~~z\leq -3 ,~H^2\geq 0.
% ~~~\mbox{for} ~~~s_A=1,~W\geq \frac{4}{L^2},
\end{eqnarray}
What is more relevant for us is the third fixed point for which 
\begin{eqnarray}
W\geq \frac{4}{L^2}, ~~~z\geq 1,~H^2\geq 0.
\end{eqnarray}
%In particular,
%\begin{eqnarray}
%F3:~~~0\leq z\leq 1, ~H^2>0, ~~~&\mbox{for}& ~~~-\frac{4}{L^2}\leq W\leq 0, \nonumber \\
%z\geq 1, ~~H^2>0, ~~~&\mbox{for}&~~~-\frac{4}{L^2}\geq W. 
%\end{eqnarray}
Some comments are in order:

\begin{itemize}

\item For $L^2W=10$, the fixed point is at $z=2$, corresponding to $H^2=1$. As we already remarked, this point is particularly interesting because is not only a point of enhanced symmetry in the bulk, but it also corresponds to the dynamical Lifshitz exponent on the brane favoured for our  observed universe and obtained  for a spectral index $n\simeq -1.6$. 

\item In a Einstein-de Sitter universe always dominated by dark matter, studied for instance in Ref. \cite{zp},  the brane dynamical Lifshitz exponent remains constant, say $\widetilde{z}=1/3$ for $z=2$. This is trivially reproduced in our set-up  taking the gauge field configuration $H^2=1$ along all the RGE flow.

\item Values of $n$ different, but in any case in the vicinity  of $n=-1.6$, correspond to  values of the bulk dynamical exponent $z$ slightly different from   $z=2$. Nevertheless all  these  points with $z\neq 2$ are  fixed points of the RGE flow.  Therefore, all of them could be used for cosmological considerations on the LSS. The price to pay, maybe not that expensive,  would be that these points are not points of enhanced symmetry.
\end{itemize}
%Let us note that at any  fixed point, the metric  reduces to the form (\ref{m01}), as it should. In particular, for $z<0$, we can write the metric at the fixed point as 
%
%\beq
%\label{m011}
%\d s^2 =- \frac{r^{2 |z|}}{L^{2 |z|}}\d t^2+\frac{L^2}{r^2}\left(2\d t \d\xi+\d\vec{x}^2+\d r^2\right). 
%\eeq
%Clearly, for $r\to 0$ the metric turns out to be 
%\beq
%\label{m012}
%\d s^2 =\frac{L^2}{r^2}\left(2\d t \d\xi+\d\vec{x}^2+\d r^2\right),
%\eeq
%which is just exact $AdS_6$ and therefore, $z=1$. In other words, starting from a negative $z$, we always end up at $z=1$ as $r\to 0$.  Similarly, for $z>1$, the geometry approaches $AdS_6$ in the other limit $r\to \infty$. Therefore, starting with a  $z>1$, we always end up at $z=1$ as $r\to \infty. $
%Let us also make a short comment on the scalar field. We have introduced the latter as a way to depart from the solution of section 4 by making $z$ running (function of $r$). Alternatively,  we could have turned on other components of the gauge field. The only such component which can be switched on consistently with the symmetries is the component $A_\xi$ along the null direction $\xi$, at the cost of more complicated field equations. .  
For the running among generic fixed points where the gauge field is not zero, the gauge dynamics is important and  the full system of equations Eqs. (\ref{eq1})-(\ref{eq5}) should be considered. In particular, we are interested in the RGE flow going from $z=2$ and back to $z=2$. This corresponds on the brane to the flow of the dynamical exponent $\widetilde{z}$ from linear to highly non-linear scales.

To find an example of such a flow we follow Ref. \cite{Liu:2015xxa} and  we write the potential for the scalar field as

\beq
V(\phi)=V_0+V_1\phi+\frac{1}{2}V_2\phi^2+\frac{1}{6}V_3\phi^3+\frac{1}{24}V_4\phi^2(\phi-\phi_0)^2,
\eeq
and similarly for the function 

\beq
W(\phi)=W_0+W_1\phi+\frac{1}{2}W_2\phi^2+\frac{1}{6}W_3\phi^3.
\eeq
Assuming a flow from $\phi= 0$  to $\phi= \phi_0$  and taking the first derivative of
$W(\phi)$  to vanish at  the fixed points, the set of conditions can be chosen as

\begin{eqnarray}
V_0&=&-\frac{10}{L_{\rm IR}^2}=-W_0,\nonumber\\
V_1&=&0=W_0,\nonumber\\
V_2\phi_0^2&=&-60\left(\frac{1}{L_{\rm UV}^2}-\frac{1}{L_{\rm IR}^2}\right)=-W_2\phi_0^2,\nonumber\\
V_3\phi_0^3&=&120\left(\frac{1}{L_{\rm UV}^2}-\frac{1}{L_{\rm IR}^2}\right)=-W_3\phi_0^3.
\end{eqnarray}
The solution of the system  cannot be found in closed form. However, its form is given in
 Fig. 2 for some choice of the parameters.  In this   example $z$ runs from $z=2$ back to $z=2$ (and correspondingly $\widetilde{z}$ runs from $\widetilde{z}=1/3$ back to $\widetilde{z}=1/3$) from the the UV to the IR (that is at larger and larger values of the scale factor).   This running therefore catches the behavior of the LSS dynamical exponent from the linear to the NL phase passing through the QNL stage, when indeed $\widetilde{z}$ is smaller and perturbations runs faster in time.
\begin{figure}[h!]
    \begin{center}
      \includegraphics[scale=.5]{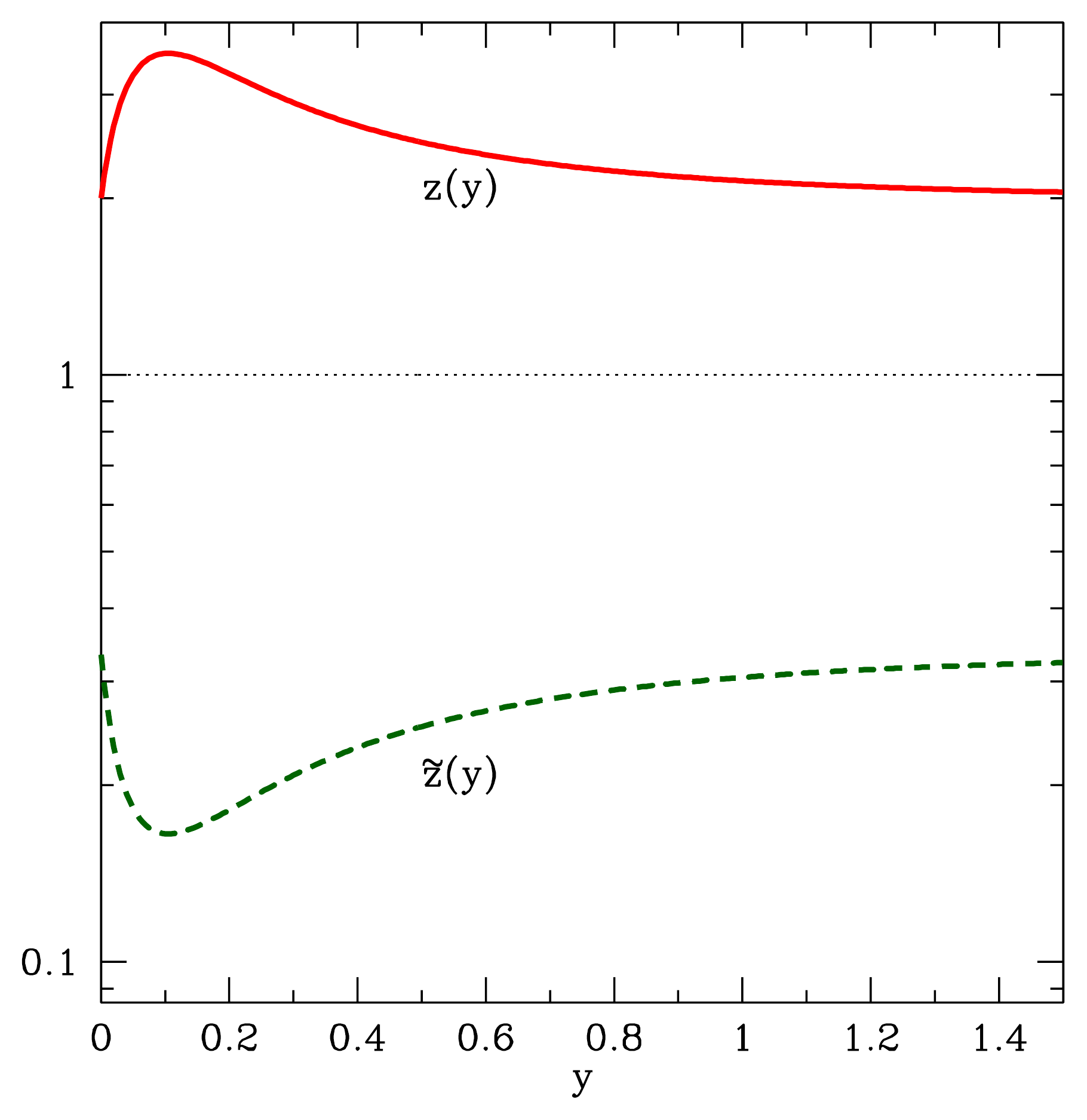}
    \end{center}
     \caption{\small A possible example of the evolution of the  exponents $z$ and $\widetilde{z}$  as a function of the coordinate $y$. They  run from $z=2$ and $\widetilde{z}=1/3$
to the same values. These two regions correspond to linear and non-linear regimes.  The values are taken to be $L_0=1$, $L_{\rm UV}=11 L_0/10 $, $V_0=-W_0=-10$, 
$V_1=-W_1=0$, $V_2=-W_2= -60(1/L_{\rm UV}^2-1)$, $V_3=-W_3= 120(1/L_{\rm UV}^2-1)$, $V_4=24/\phi_0^2$, $\phi_0=1$.
} 
\end{figure}
This running of the dynamical exponent $z$ in the bulk, corresponding to changes of the dynamical exponent  $\widetilde{z}$, in Fig. 2 is of course only  exemplary.  A more realistic  one should reproduce real data.

In the example above we have assumed that the dark matter on the brane is provided by a dominant source given by the projected five-dimensional energy density $\rho_5$. In the following we elaborate about relaxing this option and  realizing the  dark matter on the brane in terms of a dual field in the bulk, in such a way that   the symmetry properties of the bulk is  inherited on the brane.

\section{Holographic dark matter}
Comparing with the brane-world holography in the usual Randall-Sundrum setup \cite{rs}, it is well known that the symmetries of AdS space allows for a radiation fluid on the brane, a relativistic analogue of our $\rho_5$ above. This radiation dominated brane-world cosmology can however be understood from a holographic point of view, where instead of a radiation fluid added by hand on the brane, a black hole is added in the bulk, and the emergent radiation dominated universe on the brane is purely a higher dimensional geometric effect due to a black hole in the bulk. It is believed that a dual CFT, slightly broken by the brane location, describes the ``dark radiation" on the brane, and a 
bulk-brane duality exists \cite{dr1,dr2,dr3}. Heuristically, one can think that it is the Hawking radiation of the bulk black hole that is heating the brane.

Similarly it would be desirable if we could set $\rho_5=0$, and realize dark matter on the brane by slightly altering the bulk geometry instead. In this way, if the dark matter on the brane is realized in terms of a dual field in the bulk, we  expect that the fluid on the brane  inherits the symmetry properties of the bulk.

In fact, it is clear that if the bulk geometry is altered in such a way that  $L=L(R)$ and   $L(R)$ has the right functional form, the geometry will lead to an "optical illusion" of ``dark matter" on the brane. The holographic dual of dark matter, which will give the correct function form of $L(R)$, will be in our set-up simply the  scalar field $\phi$ in the bulk  we encountered in the previous section to cause the running of the bulk  dynamical exponent plus a ghost-like field $\theta$.  The addition of these scalar fields leading to dark matter on the brane is of course a perturbation of the system from the Lifshitz invariant fixed point, but again  it is an irrelevant perturbation and the system will flow immediately back towards the fixed point.

Let us therefore investigate  the issue of the holographic dark matter  and explore further the Friedmann equation in Eq. (\ref{Hb}).  We will assume  $\rho_5=0$, so that the brane has only tension $\sigma$.
Let us first provide some general considerations. The function  $L(R)$ has to decrease when going from the UV to  the  IR  (in other words from smaller to larger values of the scale factor).  From now on we take the variable $y$ to be positive. 

We write the  asymptotic behavior of the function $L(y)$  as

\beq
L(y) \simeq A+ De^{- C y},\,\,\, C>0,
\eeq
when $y\to+ \infty$. Therefore

\beq
\frac{1}{L^2}\simeq \frac{1}{A^2}\left(1-2\frac{D}{A}e^{- C y}\right).
\eeq
The function $B(y)$ is such that $B'(y)=1/L(y)$ and therefore we get

\beq
B\simeq \frac{y}{A}.
\eeq
Because of Eq.  (\ref{df}),   we know that the scale factor is related to the function $B(y)$ as $a=e^{B}$. This means that 
for $y\to+\infty$ we have  $a=e^{B}=e^{y/A}$ and

\beq
-\frac{1}{L^2}\simeq -\frac{1}{A^2}\left(1-2\frac{D}{A}a^{- C A }\right).
\eeq
From this expression we infer 

\beq
C\cdot A=+3\Rightarrow A>0,\,\,\,\,\, D>0 \,\,\,{\rm for}\,\,\, y\to +\infty.
\eeq
In other words, in order to reproduce the correct Friedmann equation the function  $L(y)$ needs to  reach a  positive  asymptote  at $y\to + \infty$ from above. This means that around the asymptotes the function 
 $L(y)$ is decreasing, as it should. This is not consistent with Eq.  (\ref{eq4})
 
 \beq
(\phi')^2=8 \frac{L'}{L^2}>0.
\eeq
 So, the behaviour of $L(y)$ is not consistent with the holographic $c$-theorem \cite{Liu:2012wf} stating that $L'$ has to be positive. In order to bypass this problem, 
we need to consider a scalar field with the wrong-sign kinetic term.  The full action becomes 
\begin{eqnarray}
 {\cal S}&=&\int \d^6 x\sqrt{-g}\bigg(R-2 V(\phi)-\frac{1}{2}(\partial\phi)^2  -\frac{1}{4}
 F_{\mu\nu}F^{\mu\nu}-\frac{1}{2}W(\phi) A_\mu A^\mu\nonumber\\
 &+&\frac{1}{2}(\partial\theta)^2 -2U(\theta)\bigg).
 \end{eqnarray} 
All  field equations are the same as before, but the following
\begin{eqnarray}
0&=&\phi'^2
-\theta'^2
-\frac{8}{L^2}L', \label{eq40bis}\\
0&=&\frac{1}{4}\phi'^2
-\frac{1}{4}\theta'^2
-V
-U
-\frac{10}{L^2}, \label{eq50bis}\\
0&=&\theta''+\frac{5}{L} \theta'+2 \partial_\theta U.\label{eq60}
\end{eqnarray}  
\begin{figure}[h!]
    \begin{center}
      \includegraphics[scale=.5]{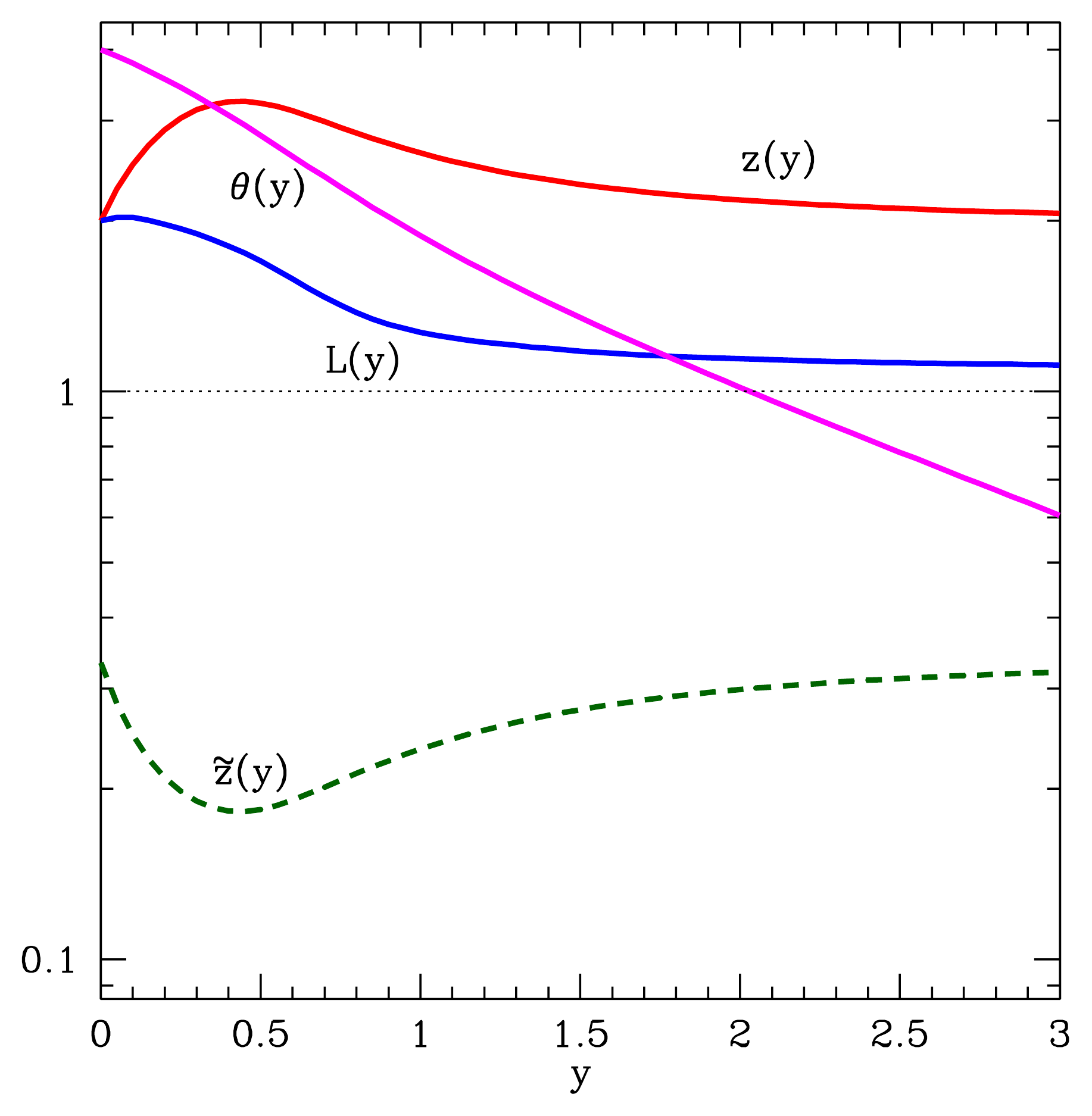}
    \end{center}
     \caption{\small A possible example of the evolution of the  exponents $z$, $\widetilde{z}$ and $L(y)$  as a function of the coordinate $y$. They  run from $z=2$ and $\widetilde{z}=1/3$
to the same values and $L(y)$ has the correct behaviour to mimic the dark matter. The values are like in Fig. 2, apart from  $L_0=2$.} 
\end{figure}
Again, analytical solutions are not available, but we show an example in Fig. 3,  where we have taken the potential for the ghost field
$\theta$ to be $U(\theta)=2/L_0^2\,\theta^2$.
 In particular, we see that the effect of the wrong-sign in the kinetic term of the $\theta$ field
makes the function $L(y)$ to decrease, as needed to reproduce the correct cosmology if we take $\rho_5=0$. In this case, taking 

\beq
\frac{1}{A^2}=\frac{1}{64 M_*^8\sigma^2}+\Lambda_{\rm obs},\,\,\, 2\frac{D}{A^3}=\frac{8\pi G_{\rm N}}{3},
\eeq
we can reproduce the usual Friedmann equation with a tiny cosmological constant $\Lambda_{\rm obs}$ plus non-relativistic dark matter. Of course, using a ghost scalar field might not be particularly appealing to some reader, but this seems to be the price to pay when abandoning the more standard beaten   path of populating the brane with dark matter from a five-dimensional $\rho_5$. On the other end, it would be interesting to further investigate the possible relation between the wrong-sign kinetic term of the field $\theta$ with the phenomenon of the gravitational instability on the brane.

\section{Criticisms, perspectives and conclusions \label{sec:conclusionsl}}
In this paper we have taken the first step towards the construction of a gravity dual of the LSS dynamics of our universe. Our considerations
are based on the realization that such   dynamics respects a Lifshitz symmetry which may manifest itself as an isometry in a six-dimensional bulk where  our universe is located as  a four-dimensional brane. 

We have shown  that the $z=2$ value of the bulk Lifshitz dynamical exponent is matched during a matter-dominated period into a brane  Lifshitz dynamical exponent which corresponds to the value observed in our universe. This is interesting as  $z=2$ is a point of enhanced symmetry. Furthermore, 
the holographic dynamical exponent  in the dual theory can perform an RGE flow between Lifshitz fixed points, going for instance from $z=2$ back to $z=2$.  We find it encouraging that the corresponding value of the dynamical exponent on the boundary theory  describes the LSS of our observed universe at large and small scales, while deviating in between. This RGE flow of the bulk dynamical exponent is reflected in a growth of the LSS in the non-linear regime with roughly the same rate it has in the the linear regime, but much slower than in the quasi non-linear regime, something that is also observed in current data. 

Let us now stress what are (some of) the incomplete aspects of our work:

\begin{itemize}

\item We have focussed more on the qualitative aspects of the construction of the gravity dual to show that the RGE flow of the bulk dynamical exponent can  explain   the running of the brane dynamical exponent observed in our universe. Ultimately, we will have to deal with the  more quantitative aspects of the problem.

\item A basic point of any gravity dual correspondence is to associate a field in the bulk 
 with the  appropriate  operator on the  boundary. While we elaborate on this step in appendix A, the ultimate goal will be to compute the boundary correlators, for instance of the dark matter density contrast, by solving the bulk gravity equations of motion with the appropriate boundary conditions. This will provide us with the boundary correlators although in a flat background. In order to get the correlators in a curved background, one has to let the brane move, thus producing the cosmological expansion in its world volume. At present, the correspondence we have put forward is at a pure level of a conjecture and 
we leave these steps 
for future work. 

\item Related to the previous point, at this stage we are  not able to make any prediction about the correlators, for instance, the power spectrum,  of the dark matter on small scales. 

\item A final point concerns the dual CFT of the gravitational theory we discussed here (and in fact for all gravity duals describing low-energy condensed matter theories). It should be stressed  that the background Lifshitz spacetime is not asymptotically locally AdS and  one cannot trivially extend standard holographic results in this case. 
%In addition, there is no even a reason to expect that there is an ordinary local quantum field theory for the holographic dual.
For the particular case of a five-dimensional  Schr\"odinger spacetime, it has been noted already in Ref. \cite{Son:2008ye} that a holographic dual should exists as the Schr\"odinger spacetime  can be obtained by a irrelevant deformation of AdS$_5$ \cite{Kostas1}. Moreover, it was argued in Ref. \cite{MMT} that the dual theory in this case is a null dipole theory, based on the fact that the Schr\"odinger spacetime can be obtained from AdS$_5$ via  TsT transformations. Similarly we expect that the dual theory of the six-dimensional  Schr\"odinger spacetime we employed here to be related to the AdS$_6$ vacuum of massive IIA theory via a chain of dualities and singular boosts. The dual theory is expected to be   a null dipole theory, arising from a null dipole deformation of a 5D large $N$ SYM theory in this case as well.

\end{itemize} 
It is clear that the next task will be  to establish this conjecture on a firmer footing and use it to explore structure formation in the non-linear regime. We leave it for future work. If successful, an unknown area in the analytical understanding of LSS  formation will be charted. In any case, the set-up provides an interesting new arena for testing the limitations of holography and gravitational duals as a tool  of studying strongly coupled field theories.

%A final point concerns the dual CFT of the gravitational theory we discussed here. It should be stressed though that as the background Lifshitz spacetime is not asymptotically locally AdS, one cannot trivially extend standard holographic results in this case. In addition, there is no even reason to expect that there is an ordinary local quantum field theory for the holograpic dual. However, for the particular case of a 5D Schr\"odinger spacetime, it has been noted already in \cite{Son:2008ye} that it can be obtained by a irrelevant deformation of $AdS_5$. Moreover, it was argued in \cite{MMT} that the dual theory in this case is a null dipole theory, based on the fact that the Schr\"odinger spacetime can be obtained from $AdS_5$ via  TsT tranformations. Similarly we expect that the dual theory of the 6D Schr\"odinger spacetime we employed here to be related to the $AdS_6$ vacuum of massive IIA theory via a chain of dualities and singular boosts. The dual theory is expected to be to be  a null dipole theory in this case as well. 
%

%\clearpage

\section*{Acknowledgments}
We thank G. D'Amico, G. Gabadze, M. Maggiore, M. Pietroni, O. Pujolas and J. Sonner for many  criticlal discussions throughout the completion of this work  and especially V. Desjacques mor many illuminating discussions about the properties of the LSS.
A.R. is supported by the Swiss National Science Foundation (SNSF), project {\sl Investigating the
Nature of Dark Matter}, project number: 200020-159223. M.S.S. is supported by the Lundbeck foundation and Villum Fonden  grant 13384.
CP3-Origins is partially funded by the Danish National Research Foundation, grant number DNRF90.

\appendix
\numberwithin{equation}{section}
%
%%%%%%%%%%%%%%%%%%%%%%%%%%%%%%%
\section{Identifying dual fields}
%%%%%%%%%%%%%%%%%%%%%%%%%%%%%%
As we already remarked, the final goal of any gravity dual of the LSS is to  find the dynamics of the relevant   boundary observables defined by the theory in the bulk   and to construct  boundary correlators (such as the one for the density contrast) as given by the value of the renormalized
bulk action for specified boundary values of the bulk fields in the Lifshitz backgrounds \cite{Son:2008ye,  Balasubramanian:2008dm,Adams:2008wt,kak,RR0,Kostas,L1,L2,L3,L4,L5,L6}. In this appendix we take a first step by  identifying  which the bulk dual fields corresponds to  the  boundary fields.

A convenient way for our purposes is to perturb the coordinate basis $\d x^M$ according to
\beq
\d x^M\to \d x^M+{h^M}_N\, \d x^N, 
\eeq
in such a way that the null structure of spacetime is preserved, that is $\partial_\xi$ remains a null vector. Then, the possible
perturbations are
\begin{eqnarray}
\d t&\to& \d t+{h^0}_0 \d t+{h^0}_i\,\d x^i,\nonumber\\
 \d\xi&\to& \d\xi+{h^\xi}_0\, \d t+{h^\xi}_\xi \,\d\xi+{h^\xi}_i\,\d x^i,
\nonumber \\
\d x^i&\to& \d x^i+{h^i}_0\,\d t+{h^i}_j \,\d x^j.
\end{eqnarray}
The above change in the coordinate basis, give rise to a corresponding metric perturbation
\begin{eqnarray}
\d s^{2}&=&-\frac{L^{2z}}{r^{2z}} \Big{[}(1\!+\!2{h^0}_0)\d t^2\!+\!2 {h^0}_i \d t \d x^i\Big{]}\!+\!2\frac{L^2}{r^2}
\left[\left(1\!+\!\frac{1}{2} {h^0}_0\right)\d t \d\xi\!+\!{h^\xi}_0 \d t^2\!+\!{h^0}_i \d x^i \d\xi\!+\!{h^\xi}_i \d x^i \d t\right]\nonumber \\
&+&\frac{L^2}{r^2}\left[\Big{(}\delta_{ij}+h_{ij} \Big{)}\d x^i \d x_j +\d r^2\right] .\label{gper}
\end{eqnarray}
 Note that, since there is an O(3) symmetry of the metric, 
 we may decompose the metric perturbations according to their transformation properties under rotations.
Thus, there should exist scalar, vector and tensor metric perturbations. Under this decomposition we have
\begin{itemize}
\item scalar perturbations: (${h^0}_0, ~ {h^0}_\xi, ~ {h^\xi}_\xi$),
\item vector perturbations: (${h^0}_i, ~ {h^\xi}_i, ~ {h^i}_0$),
\item tensor perturbations:  (${h^i}_j$).
\end{itemize}
 %We will consider here only scalar perturbations
%since their boundary values are coupled to the energy density of the boundary theory.
If we assume that the perturbations ${h^M}_N$ preserve Lifshitz scaling,  they should transform as
\begin{eqnarray}
{h^0}_0&\to&  {h^0}_0,\nonumber\\
{h^0}_i&\to& \lambda^{z-1} {h^0}_i,\nonumber \\
 {h^\xi}_\xi&\to& {h^\xi}_\xi'={h^\xi}_\xi,\nonumber\\
{h^0}_\xi&\to& \lambda^{2(z-1)}{h^0}_\xi,\nonumber \\
 {h^\xi}_i&\to&\lambda^{1-z}{h^\xi}_i, \nonumber\\
 {h^i}_0&\to&\lambda^{1-z}{h^i}_0,\nonumber \\
{h^i}_j &\to& {h^i}_j.
\end{eqnarray}
These metric perturbations will couple to the components of the energy-momentum tensor of the boundary theory. Equivalently, the energy-momentum tensor in this case \cite{RR0}, which is not-symmetric due to lack of Lorentz invariance,  may be coupled  to the vielbeins \cite{Kostas}.  Let us recall that
in a non-relativistic theory described by a non-relativistic Lagrangian ${\cal{L}}(q_a,\dot{q}_a)$, the components of the energy-momentum tensor
are
\begin{eqnarray}
{\cal{E}}&=&T^{00}=\sum_a p_a\dot{q}_a-{\cal{L}}, \nonumber \\
 S^i&=&T^{i0}=\sum_a \dot{q}_a\frac{\partial {\cal{L}}}{\partial\partial_iq_a},\nonumber \\
{\cal{P}}_i&=&T^{0i}=-\sum_a\frac{\partial {\cal{L}}}{\partial\dot{q}_a} \partial_i q_a,\nonumber \\
{T^{i}}_j&=&-\sum_a \frac{\partial {\cal{L}}}{\partial\partial_iq_a} \partial_j q_a+\delta^i_j {\cal{L}}.
\end{eqnarray}
It is straightforward to see the scaling properties of the components of the energy-momentum tensor under Lifshitz scalings 
\begin{eqnarray}
{\cal{E}}&\to& \lambda^{-3-z} {\cal{E}}, \nonumber \\
 S^i&\to& \lambda^{-2(1+z)}S^i, \nonumber \\
{\cal{P}}_i&\to& \lambda^{-4}{\cal{P}}_i, \\
 {T^{i}}_j&\to& \lambda^{-3-z}{T^{i}}_j.
\end{eqnarray}
Then the scale-invariant coupling of the metric perturbations of bulk to the components of the energy-momentum
tensor of the boundary theory are then uniquely determined to be
\beq
S=\int \d^4x\left({h^0}_0{\cal{E}}+{h^0}_i S^i+{h^\xi}_0\rho+{h^i}_0 {\cal{P}}_i+{h^\xi}_i {\cal{P}}_i+{h^i}_j{T^j}_i\right).
\eeq
This identifies $h^\xi_0$ as the  dual field which will provide correlators of the perturbed dark matter energy density $\rho$. 

\end{document}